\documentstyle[12pt]{article}

\newcommand{\be}{\begin{equation}}
\newcommand{\ee}{\end{equation}}

\newcommand{\bt}{\beta}

\newcommand{\ep}{\varepsilon}
\newcommand{\al}{\alpha}
\newcommand{\ra}{\rightarrow}

\newcommand{\gm}{\gamma}
\newcommand{\om}{\omega}

\newcommand{\lbd}{\lambda}

\begin{document}

\begin{center}

{\Large{\bf Extrapolation and interpolation of asymptotic series 
by self-similar approximants} \\ [5mm]
V.I. Yukalov$^{a*}$, E.P. Yukalova$^b$, and S. Gluzman$^c$} \\ [3mm]

{\it $^a$Bogolubov Laboratory of Theoretical Physics, \\
Joint Institute for Nuclear Research, Dubna 141980, Russia, \\ [2mm]

$^b$Laboratory of Information Technologies, \\
Joint Institute for Nuclear Research, Dubna 141980, Russia, \\ [2mm]

$^c$Generation 5 Mathematical Technologies Inc., \\
Corporate Headquaters, 515 Consumer Road, Toronto, 
Ontario M2J 4Z2, Canada}

\end{center}

\vskip 1cm

{\parindent=0pt
$^*$Corresponding author

\vskip 1cm
{\bf Mailing address}:  

\vskip 2mm
Bogolubov Laboratory of Theoretical Physics,

Joint Institute for Nuclear Research,

Dubna 141980, Russia

\vskip 5mm
{\bf Telephone number}: 7(496)216-3947

\vskip 1mm
{\bf Fax number}:       7(496)216-5084

\vskip 1mm
{\bf E-mail address}:   yukalov@theor.jinr.ru

}

\newpage

\begin{abstract}

The problem of extrapolation and interpolation of asymptotic 
series is considered. Several new variants of improving the 
accuracy of the self-similar approximants are suggested. The 
methods are illustrated by examples typical of chemical physics, 
when one is interested in finding the equation of state for a 
strongly interacting system. A special attention is payed to the 
study of the basic properties of fluctuating fluid membranes. It 
is shown that these properties can be well described by means of 
the method of self-similar approximants. For this purpose, the 
method has been generalized in order to give accurate predictions 
at infinity for a function, whose behavior is known only at the 
region of its variable close to zero. The obtained results for 
fluctuating fluid membranes are in good agreement with the known 
numerical data.

\end{abstract}

\vskip 2cm

{\parindent=0pt
{\bf KEY WORDS}: asymptotic series, summation of series, methods 
of extrapolation and interpolation, self-similar approximation 
theory, equations of state, fluctuating fluid membranes

}

\newpage

\section{Introduction}

Asymptotic series are ubiquitous, arising in practically all 
realistic problems that do not allow for exact solutions but 
require the use of some kind of perturbation theory. The latter 
assumes that there exists a small parameter, such that the 
observable quantities of interest can be represented as series 
in powers of this asymptotically small parameter. However, in 
applications, this parameter is not negligibly small, but takes 
finite values. In the majority of cases, the perturbative series 
are divergent for the parameter finite values, corresponding to 
realistic systems. This is why the problem of extrapolating 
asymptotic series is of great importance. The problem poses the 
question what are the most accurate ways of extending the validity 
of the series, obtained for asymptotically small parameters, to 
the finite values of the latter. Moreover, in some cases it is 
necessary to extend perturbative results to the extreme limit, 
where the parameter tends to infinity.

The most popular techniques of extrapolating asymptotic series 
are based on the Pad\'e approximants [1]. However, these have 
several well known shortcomings. First of all, they are not 
uniquely defined. For a series of a given order, there is a whole 
table of many Pad\'e approximants, but there is no general recipe 
that would  advise which of them should be preferred. The conclusion 
could be made if all these approximants from the table would be 
close to each other. Then their dispersion would define their 
accuracy. This, unfortunately, is a very rare case, since a rather 
standard situation is when the results of the approximants from the 
table are widely scattered, making it difficult to judge on their 
accuracy. Even worse, among the approximants from the table, there 
very often appear those that contain unreasonable poles. Then the 
accuracy of such approximants, strictly speaking, is not defined 
at all. One can, of course, neglect the approximants with the poles, 
calling them outliers. But, as is clear, this is a quite subjective 
procedure, since it may happen that the arising poles could have 
meaning, as is the usual case for critical phenomena. Arbitrarily 
rejecting some of the approximants, while keeping others, renders 
the whole procedure not well defined. Then one is not able to 
ascribe any accuracy to the obtained results. The most difficult 
situation is when one needs to extrapolate the perturbative series 
to the infinite value of the expansion parameter. In such a case, 
the Pad\'e approximants cannot be used at all. In order to be 
specific, showing that this necessity of extrapolating the parameter 
to infinity does happen in realistic cases, we can mention the 
problem of calculating the pressure of fluctuating fluid membranes.

Different types of membranes are rather frequent structures in 
chemical and biological systems [2--6]. The membrane thermal 
fluctuations between two hard walls are often described by field 
theory. An important class of membranes are fluid membranes, whose 
constituent molecules are able to move within them. The fluctuations 
are controlled by their bending rigidity.

When applying field theory to the description of membranes, 
one encounters the following problem. First, to proceed in 
calculations, one replaces the hard walls by a smooth potential of 
a finite stiffness, which, in dimensionless units, can be denoted 
as $g$. Then, one is able to proceed by invoking perturbation theory 
in powers of $g$. However, to return back to the sought case of 
rigid walls, one needs to set $g\ra\infty$. Thus, the problem arises 
how from an expansion in powers of small $g\ra 0$ one could extract 
information on the quantities of interest for large $g\ra\infty$? 
The standard resummation techniques, such as Borel or Pad\'e [6] 
could help in extending asymptotic series in small $g\ra 0$ to 
finite values of $g$, but these techniques are not applicable for 
the limit of $g\ra\infty$.

The problem of extending a function $f(g)$, which is known only 
for asymptotically small $g\ra 0$, to the whole region of $g$, 
including the limit $g\ra\infty$, can be solved by the {\it 
optimized perturbation theory}, advanced in Ref. [7]. This theory 
has been successfully applied to a variety of problems, as can be 
inferred from the review-type articles [8,9] and references therein. 
The pivotal idea of the optimized perturbation theory [7] is to 
introduce in the calculational process control functions defined 
by optimization conditions. As a result, a function $f(g)$, known 
only for small $g\ra 0$, can be extrapolated to the whole region of 
$g$, including the limit $g\ra\infty$. There exist three main ways 
of introducing control functions. One way is to include in the 
initial approximation trial parameters that are transformed, by 
means of the optimization conditions, into control functions at 
each step of perturbation theory. Another way is to introduce 
control functions in the process of accomplishing a perturbative 
or iterative scheme, for instance, by defining the cutoffs of 
integrals or introducing regularization masses, which are then 
to be transformed into control functions. The third way is to 
derive, first, an asymptotic series in powers of a small parameter 
or variable and then to reorganize the derived series by means of 
a change of variables, with control functions included in this 
variable transformation. Various examples of introducing control 
functions can be found in literature [10--30] (see also the review 
articles [8,9]).

The optimized perturbation theory has been applied to fluid 
membranes in several papers. The most accurate results have been 
obtained by Kastening [31], whose calculations are based on the 
sixth-order perturbation theory, with introducing control functions 
by means of the Kleinert change of variables [32]. This method 
requires rather heavy numerical calculations. Also, the Kastening 
result [31] slightly deviates from the Monte Carlo [33] simulations 
for this problem.

The aim of the present paper is to reconsider the general problem 
of extrapolating asymptotic series in order to find an accurate 
and simple way of accomplishing such an extrapolation. We also 
consider the case when the behavior at infinity is known, which 
then becomes the problem of an accurate interpolation. We pay a 
special attention to developing simple methods for obtaining the 
limit $f(\infty)$ at $g\ra\infty$ from the asymptotic expansion 
of $f(g)$ at $g\ra 0$. To  illustrate the methods, we apply them 
to several problems, which yield the series, whose mathematical 
structure is typical for many real situations in chemical physics. 
At the end, we apply these methods to considering the problem of 
fluid fluctuating membranes.

The method we aim at developing is based on the self-similar
approximation theory [34--41] in the variant involving the 
self-similar root approximants [42--44] and self-similar factor 
approximants [45--49]. This approach has been applied to variety
of problems, providing high accuracy and at the same time being 
quite simple [42--49]. It was shown to be essentially more accurate 
than the use of Pad\'e approximants [42--52]. However, in some 
cases, when the self-similar approximants were directly applied 
for solving the problems involving the limit of $g\ra\infty$, the 
results were not satisfactory. Our aim now is to generalize the 
method of self-similar approximants so that it would provide good 
accuracy for the limiting value of $f(\infty)$ at $g\ra\infty$. 
Then we apply the method to solving several problems requiring the 
construction of the equations of state for strongly interacting 
systems. Among them, we study the  problem of fluctuating fluid 
membranes and demonstrate that the newly developed methods provide 
good accuracy by comparing our results with the known numerical 
calculations.

\section{Factor and root approximants}

First, we need to recall the method of self-similar factor and 
root approximants [42--52], which we are going to improve. Suppose, 
we are interested in the behavior of a real function $f(g)$ of a 
real variable $g$. And let us assume that this function is defined 
by so complicated equations that the sole thing we are able to find 
is the property of the function at asymptotically small $g$, where
\be
\label{1}
f(g) \simeq f_k(g) \qquad (g\ra 0)
\ee
is approximated by the series
\be
\label{2}
f_k(g) =  f_0(g) \sum_{n=0}^k a_n g^n \; ,
\ee
in which $f_0(g)$ is a form that cannot be expanded in powers 
of $g$. Without the loss of generality, we can set $a_0=1$, since 
any $a_0$ not equal to one can be incorporated into $f_0(g)$.

Series (2), which are valid for $g\ra 0$, can be extrapolated to the
region $g>0$ by means of the self-similar factor approximants [45--49]
having the form
\be
\label{3}
f_k^*(g) =  f_0(g) \prod_{i=1}^{N_k}\; ( 1 + A_i g)^{n_i} \; ,
\ee
where
\begin{eqnarray}
\label{4}
N_k = \left \{ \begin{array}{ll}
k/2 \; , & \; \; k=2,4,\ldots \\
(k+1)/2 \; , & \; \; k=3,5,\ldots
\end{array} \right.
\end{eqnarray}
The parameters $A_i$ and $n_i$ are defined by the {\it re-expansion
procedure}, when the $k$-th order approximant (3) is expanded in 
powers of $g$ up to the $k$-th order as
\be
\label{5}
f_k^*(g) \simeq f_0(g) \sum_{n=0}^k a_n^* g^n \; ,
\ee
where $a_n^*=a_n^*(\{ A_i\},\{ n_i\})$. Then expansions (2) and (5) 
are compared, with equating the same-order terms
\be
\label{6}
a_n^*(\{ A_i\},\{ n_i\}) = a_n \; .
\ee
This way is also called the accuracy-through-order procedure.

In order to give an explicit representation of Eqs. (6), it is 
convenient to equate the logarithms
\be
\label{7}
\ln  f_k^*(g) \simeq \ln  f_k(g) \qquad (g\ra 0) \; ,
\ee
which, taking into account form (3), yields
\be
\label{8}
\sum_{i=1}^{N_k} n_i \ln ( 1 + A_i g) \simeq
\ln \sum_{m=0}^k  a_m g^m \; .
\ee
Expanding here
$$
\ln ( 1 + A_i g) = \sum_{m=1}^\infty \; \frac{(-1)^{m-1}}{m} \;
( A_i g)^m \; ,
$$
we come to the equations
\be
\label{9}
\sum_{i=1}^{N_k}\; n_i A_i^n =  B_n \qquad ( n =1,2,\ldots, k) \; ,
\ee
in which
\be
\label{10}
B_n = \frac{(-1)^{n-1}}{(n-1)!} \; \lim_{g\ra 0} \;
\frac{d^n}{dg^n}\; \ln \left ( \sum_{m=0}^n a_m g^m \right ) \; .
\ee
The system of equations (9) contains $k$ equations. When $k$ is 
even, system (9) defines all $k/2$ parameters $A_i$ and $k/2$ 
parameters $n_i$. When $k$ is odd, then there are $k+1$ parameters, 
$(k+1)/2$ parameters  $A_i$, and $(k+1)/2$ parameters $n_i$. Then 
the system (9) is complemented by the condition $A_1=1$ following 
from the scaling arguments [47,48]. Thus, the re-expansion procedure 
completely defines all parameters of the self-similar approximant 
(3). It may happen, though it is a rather rare case, that Eqs. (9), 
for some order $k$, do not have solutions. Then one just needs to 
proceed to the higher orders of the series. But it is important to stress that for each given order $k$ the factor approximants are 
uniquely defined. So, when Eqs. (9) possess solutions, these 
solutions are unique.

The described above method allows for the extrapolation of a series 
for a small $g \ra 0$ to the finite values of $g$. In some cases, 
there can exist additional information on the behavior of the function 
at asymptotically large $g$,
\be
\label{11}
f(g) \simeq f_p(g) \qquad (g \ra \infty) \; ,
\ee
so that
\be
\label{12}
f_p(g) = \sum_{j=1}^p b_j g^{\al_j} \qquad 
(\al_j > \al_{j+1} ) \; ,
\ee
with the powers $\alpha_j$ in the descending order. In that case, 
we have the problem of interpolation between small $g \ra 0$ and 
large $g \ra \infty$. Suppose that $f_0(g)$ at large $g$ behaves 
as
\be
\label{13}
f_0(g) \simeq A g^\al \qquad ( g \ra \infty)   \; .
\ee
Then, in order that the factor approximant (3) would satisfy the 
limiting form (12), we need to set
\be
\label{14}
A \prod_{i=1}^{N_k} A_i^{n_i} = b_1 \qquad 
\al + \sum_{i=1}^{N_k} n_i = \al_1  \; .
\ee
This type of the interpolating factor approximant, employing $k$ 
terms from the small-variable expansion and one limiting term from 
the large-variable behavior, will be denoted as $f_{k+1}^*(g)$. In 
the following sections, we develop alternative methods of 
interpolation, improving the accuracy of the factor approximants

The problem of interpolation can also be conveniently solved by
involving the self-similar root approximants [42-44], having the
form
\be
\label{15}
R_p^*(g) = f_0(g) \left ( \ldots \left ( \left ( 
1 + A_1 g\right )^{n_1} + A_2 g^2 \right )^{n_2} + \ldots +
A_pg^p \right )^{n_p} \; .
\ee
The parameters $A_i$ and $n_i$ are defined by the large-variable
expansion (12). Again, it is important to emphasize that this
definition is unique [9]. If, instead, we try to define the
parameters of a $k$-order root approximant by the
accuracy-through-order procedure, expanding Eq. (15) in powers 
of $g$ and equating the resulting expansion with Eq. (2), then we
confront the problem of nonuniqueness of solutions for the sought
parameters [50]. In the following sections, we shall suggest a way
of solving this problem. It is worth noting that the regions 
of small $g$ and large $g$ can be easily interchanged by the change 
of the variable $g$ to $1/g$.

\section {Problem of self-similar interpolation}

One of the well known difficulties in dealing with asymptotic series
occurs when the number of their terms is small, which does not allow
one to construct higher order approximants. In the present section,
we suggest a way of overcoming this difficulty. The method, we
advance,  reminds the learning algorithms used in statistical
learning [53]. The idea is as follows. Suppose we have $k$ terms
$a_k$ of the small-variable expansion and the limiting form of the
large-variable behavior. Interpolating from the right to left, that
is, considering the variable $1/g$, we construct the corresponding
self-similar approximant, say, the root approximant $R^*_{k}(g)$.
Then we expand the latter in powers of $g$ up to the $(k+1)$-order,
obtaining an additional term $a^*_{k+1}$. Using the new expansion,
we define the approximant $R^*_{k+1}(g)$. Expanding this up to the
$(k+2)$-order, we find the $(k+2)$-order term $a^*_{k+2}$. Then, 
we construct the approximant $R^*_{k+2}(g)$, and so on. Thus, 
each approximant defines the higher-order term of the small-$g$
expansion. Of course, this procedure can work only when the sought
function pertains to the class of monotonic functions and the
interpolation problem is considered. Below, we illustrate the method
by examples where one is interested in finding the equations of
state.

\subsection {Fr\"ohlich optical polaron}

Let us consider the problem of the optical polaron, being interested 
in finding its energy $e(g)$ as a function of the coupling parameter 
$g$. It is common to employ the dimensionless notations for these 
quantities, which we use in what follows. The small-$g$ expansion 
and the large-$g$ limit can be found in the review article [54]. For 
the small-$g$ expansion, one has
\be
\label{16}
e(g) \simeq a_1 g + a_2 g^2 + a_3 g^3 \qquad ( g\ra 0 )  \;   ,
\ee
with the coefficients
$$
a_1 = -1 \; , \qquad a_2 = -1.591962\times 10^{-2} \; ,
\qquad a_3 = -0.806070\times 10^{-3} \; ,
$$
While the large-$g$ behavior is given by the Miyake limit
\be
\label{17}
e(g) \simeq B g^2 + O(1) \qquad (g \ra \infty) \; ,
\ee
where $B = -0.108513$.

Following the procedure described above, we derive the coefficients
$$
a_4^* = -5.014168 \times 10^{-5} \; , \qquad
a_5^* = -3.312472 \times 10^{-6} \; .
$$
Using these and interpolating from the right to left, with the 
variable $1/g^2$, we construct the root approximant
\be
\label{18}
R_4^*(g) = Bg^2 \left ( \left ( \left ( \left ( 1 + 
\frac{A_1}{g^2} \right )^{n_1} + \frac{A_2}{g^4} \right )^{n_2}
+ \frac{A_3}{g^6} \right )^{n_3} + \frac{A_4}{g^8} 
\right )^{n_4} \; ,
\ee
where
$$
   A_1 = 64.163254,  \qquad A_2 = 7.001856 \times 10^3,
$$
$$
   A_3 = 7.026125 \times 10^5, \qquad A_4 = 5.201706 \times 10^7,
$$
$$
   n_1 = \frac{3}{2}, \qquad n_2 = \frac{5}{4}, \qquad
   n_3 = \frac{7}{6}, \qquad n_4 = \frac{1}{8},
$$
and the root approximant
\be
\label{19}
R_5^*(g) = Bg^2 \left ( \left ( \left ( \left ( \left ( 1 + 
\frac{A_1}{g^2} \right )^{n_1} + \frac{A_2}{g^4} \right )^{n_2}
+ \frac{A_3}{g^6} \right )^{n_3} + \frac{A_4}{g^8} 
\right )^{n_4} + \frac{A_5}{g^{10}} \right )^{n_5}  \; ,
\ee
with the coefficients
$$
   A_1 = 68.38553, \qquad A_2 = 7.742967 \times 10^3, \qquad
   A_3 = 8.213401 \times 10^5, 
$$
$$
A_4 = 7.313112 \times 10^7, \qquad A_5 = 4.417553 \times 10^9,
$$
$$
   n_1 = \frac{3}{2}, \qquad n_2 = \frac{5}{4}, \qquad
   n_3 = \frac{7}{6}, \qquad n_4 = \frac{9}{8},
\qquad  n_5 = \frac{1}{10}.
$$

The accuracy of these approximants can be checked by comparing 
them with the results of the Monte Carlo simulations [54] 
accomplished for the region of $g \in [1,15]$. In all this region, 
the approximant (19) has the percentage error less than $1\%$. The
maximal error of $-1\%$ occurs at $g=10$, where this error is
comparable with the Feynman variational calculations [55]. But for
all other values of $g$ in the considered interval, the accuracy of
approximant (19) is better than the Feynman results. Comparing the
accuracy of the approximants $R^*_3(g)$, $R^*_4(g)$, and $R^*_5(g)$,
we observe numerical convergence. For instance, the maximal
percentage error of $R^*_3(g)$  is $1.5\%$ at $g=10$.

\subsection{One-dimensional Bose gas}

Let us now consider the ground-state energy $e(g)$ of the 
Lieb-Liniger model [56] as a function of the coupling parameter $g$,
again using dimensionless units. The weak-coupling expansion can be
written [57,58] as
\be
\label{20}
e(g) \simeq g + a_3 g^{3/2} + a_4 g^2 + a_5 g^{5/2}  \; ,
\ee
with the coefficients
$$
   a_3 = - 0.424413, \qquad a_4 = 0.065352, \qquad
a_5 = - 0.017201.
$$
For strong coupling, we have the Tonks-Girardeau limit
\be
\label{21}
e(g) \simeq \frac{\pi^2}{3} + O\left ( \frac{1}{g} \right )
\qquad ( g \ra \infty) \; .
\ee
Following the procedure, described at the beginning of this section,
we find
$$
   a^*_6 = 5.153629 \times 10^{-3}.
$$
Then we construct the root approximants of different orders, 
interpolating from the right to left, with the variable $1/g$, 
and compare their accuracy with numerical data [58]. The 
approximant $R^*_5(g)$ has the maximal, with respect to the whole 
range of $g \in [0, \infty)$, error of $3.4\%$ at $g = 6$. The 
maximal error of $R^*_6(g)$ is $1.75\%$ at $g = 10$. The best 
approximant is obtained using the coefficients $a_3$ and $a_4$ 
of expansion (20) and the three terms of the strong coupling 
limit that can be written [57,58] as
\be
\label{22}
e(g) \simeq \frac{\pi^2}{3} \left ( 1 \; - \; \frac{4}{g} + 
\frac{12}{g^2} \right )  \qquad ( g \ra \infty ) \; .
\ee
The corresponding root approximant is
\be
\label{23}
R_{4+3}^*(g) = \frac{\pi^2}{3} \left ( \left ( \left ( 
\left ( \left ( 1 + \frac{A_1}{g} \right )^{n_1} + 
\frac{A_2}{g^2} \right )^{n_2} + \frac{A_3}{g^3} 
\right )^{n_3} + \frac{A_4}{g^4} \right )^{n_4} + 
\frac{A_5}{g^5} \right )^{n_5}\; ,
\ee
where
$$
   A_1 = 8.126984, \qquad A_2 = 37.345427, \qquad A_3 = 164.914098,
$$
$$
   A_4 = 388.171278, \qquad A_5 = 385.382911,
$$
$$
   n_1 = \frac{3}{2} \; , \qquad n_2 = \frac{5}{4} \; , 
\qquad  n_3 = \frac{7}{6} \; , \qquad
   n_4 = \frac{9}{8} \; , \qquad n_5 = \frac{1}{5} \; .
$$
This aproximant (23) provides a very high accuracy in the whole 
range of $g \in [0, \infty)$, having the maximal error of only 
$0.023\%$ at $g = 6$. Therefore, expression (23) can be employed 
as an analytical representation for the ground-state energy of 
the Lieb-Liniger gas.

\subsection{Diluted Fermi gas}

Let us now turn to defining the ground-state energy of the spin
$1/2$ Fermi gas with attractive interactions, corresponding to the
negative scattering length $a_s$. The ground-state energy $e(g)$ 
can be written [59] as an asymptotic expansion in powers of the
dimensionless parameter $g = |k_F a_s|$, where $k_F$ is the Fermi
wave vector,
\be
\label{24}
e(g) \simeq a_0 + a_1 g + a_2 g^2 + a_3 g^3 + a_4 g^4 \;  .
\ee
Here $g \ra 0$ and
$$
   a_0 = \frac{3}{10}, \qquad a_1 = - \frac{1}{3\pi}, 
\qquad a_2 = 0.055661,
$$
$$
   a_3 = - 0.00914, \qquad a_4 = - 0.018604.
$$
In the unitary limit [60,61], when $g \ra \infty$, numerical 
calculations [62] yield
\be
\label{25}
\lim_{g\ra\infty} e(g) = 0.132 \; .
\ee

We construct the self-similar approximants of different orders and 
compare their accuracy with Monte Carlo simulations [63].The factor 
approximant $f^*_{3+1}(g)$ turns out to be analogous to the diagonal 
[2/2] Pad\'e approximant [64]. However, the root approximant
$R^*_3(g)$ is essentially more accurate. The factor approximant
$f^*_{4+1}(g)$ displays the same accuracy as $R^*_3(g)$. The maximal 
percentage error of the latter two approximants in the interval of
$g \in [0,5]$ is only about $0.2\%$.

\section{Problem of self-similar extrapolation}

The problem of extrapolation of asymptotic series is much more
difficult than that of their interpolation. In the latter case, the
large-variable limit is given, while in the former case, this limit
is not known. And often, it is exactly the limiting behavior at
large variable, which is of the most interest. In the present
section, we suggest new variants of constructing the self-similar
approximants in the extrapolation problem and illustrate these
methods by the model whose mathematical structure is typical of the
variety of physical and chemical systems.

\subsection{Iterated root approximants}

The extrapolation of asymptotic series can be done by means of the
self-similar factor approximants. But, as is mentioned in Sec.2, 
if we try to accomplish the extrapolation by using the root
approximants, we encounter the problem of nonuniqueness of defining
their parameters by the accuracy-through-order procedure. To
overcome this problem, we suggest to use the iteration method, by
keeping the fixed lower-order parameters when constructing the
higher-order approximants. Then all parameters of the root
approximants can be uniquely defined.

To be concrete, let us consider the anharmonic-oscillator model 
with the Hamiltonian
\be
\label{26}
H = -\; \frac{1}{2} \; \frac{d^2}{dx^2} + \frac{1}{2}\; x^2 +
g x^4 \; ,
\ee
where $x \in (- \infty, \infty)$ and the dimensionless coupling 
parameter is positive, being in the region $g \in [0,\infty)$. The 
weak-coupling expansion for the ground-state energy [65] reads as
\be
\label{27}
e(g) \simeq a_0 + a_1 g + a_2 g^2 + a_3 g^3 + a_4 g^4 +
a_5 g^5 + a_6 g^6 + a_7 g^7 \; ,
\ee
with the coefficients
$$
   a_0 = \frac{1}{2}, \qquad a_1 = \frac{3}{4}, \qquad 
a_2 = - 2.625,
$$
$$
   a_3 = 20.8125, \qquad a_4 = - 241.2890625, \qquad 
a_5 = 3580.98046875,
$$
$$
   a_6 = - 63982.8134766, \qquad a_7 = 1329733.72705.
$$
Our aim is to extrapolate the weak-coupling expansion (27), valid 
for asymptoticaly small $g \ra 0$, to the region of finite values 
of $g$. And we shall pay a special attention to the behavior of the 
extrapolated energy at large $g \ra \infty$, comparing it with the 
known asymptotic form
\be
\label{28}
e(g) \simeq 0.667986 g^{1/3} \qquad
(g\ra \infty) \; .
\ee
The iteration method for constructing the uniquely defined root 
approximants is elucidated in the following forms extrapolating 
the asymptotic expansion (27) to finite values of $g$. We start 
with the lowest approximant
\be
\label{29}
R_2^*(g) = \frac{1}{2} (1 + A_1 g)^{n_1} \; ,
\ee
in which
$$
   A_1 = 8.5, \qquad n_1 = 0.176 \; .
$$
The next-order approximant is
\be
\label{30}
R_4^*(g) = \frac{1}{2} \left ( ( 1 + A_1 g)^{n_2} + A_2 g^3
\right )^{n_3} \; ,
\ee
with the same $A_1$ and
$$
   A_2 = 227.719, \qquad n_2 = \frac{n_1}{n_3} =2.771,
\qquad  n_3 = 0.064.
$$
The next approximant
\be
\label{31}
R_6^*(g) = \frac{1}{2} \left ( \left ( ( 1 + A_1 g)^{n_2} + A_2 g^3
\right )^{n_4} + A_3 g^5 \right )^{n_5}\; ,
\ee
contains the same $A_1$, $n_2$, and $A_2$, with
$$
   A_3 = 3.827 \times 10^4, \qquad n_4 = \frac{n_3}{n_5} = 2.001, 
\qquad n_5 = 0.032.
$$
We shall compare the strong-coupling behavior of these root
approximants with the exact limiting form (28) and with the factor
approximant
\be
\label{32}
f_6^*(g) = \frac{1}{2} ( 1 + B_1 g)^{m_1} ( 1 + B_2 g)^{m_2}
(1 + B_3 g)^{m_3} \; ,
\ee
in which
$$
   B_1 = 26.74018, \qquad B_2 = 12.46882, \qquad 
B_3 = 3.83804,
$$
$$
   m_1 = 1.80165 \times 10^{-3}, \qquad m_2 = 0.05473, \qquad 
m_3 = 0.20047.
$$

The strong-coupling behavior of the root approximants (29) to (31)
is
\be
\label{33}
R_2^*(g) \simeq \frac{1}{2} (A_1 g)^{n_1} \; , \qquad
R_4^*(g) \simeq \frac{1}{2} \left ( A_2 g^3 \right )^{n_3} \; , 
\qquad R_6^*(g) \simeq \frac{1}{2} \left ( A_2 g^3 
\right )^{n_4n_5} \; .
\ee
While the factor approximant (32), as $g \ra \infty$, gives
\be
\label{34}
f_6^*(g) \simeq \frac{1}{2} \; B_1^{m_1}\; B_2^{m_2}\; 
B_3^{m_3} \; g^{m_1 + m_2 + m_3} \; .
\ee
Substituting here the corresponding values of the parameters yields
$$
R_2^*(g) \simeq 0.728698 g^{0.176} \; , \qquad
R_4^*(g) \simeq 0.707691 g^{0.192} \; ,
$$
\be
\label{35}
R_6^*(g) \simeq 0.707814 g^{0.192} \; , \qquad
f_6^*(g) \simeq 0.756157 g^{0.257}  \; .
\ee
Comparing these expressions with the exact asymptotic form 
(28), we see that the amplitudes of the root approximants 
provide slightly better extrapolation than the factor approximant, 
however this difference is not essential, all amplitudes being 
defined with an error of about $10\%$. In many cases, the most 
important quantity that is required to be found from the 
extrapolation procedure is the power of $g$ in the limit of 
$g \ra \infty$. Equation (35) shows that the best extrapolation 
of the power is provided by the factor approximant, whose error 
is about $20\%$, while the root approximants have a twice larger 
error. The accuracy can be improved by defining the higher-order 
approximants.

In the present section we demonstrated the application of 
the method of iterated root appproximants to the problem of 
calculating the ground-state energy of the anharmonic oscillator. 
This method works well for other problems too. For example, we 
have constructed the iterated root approximants for the problem 
of the one-dimensional Bose gas of Sec. 3.2. The root approximant 
$R_6^*(g)$, extrapolated from the left to right, gives the energy 
$e(\infty)$, as $t\ra\infty$, equal to $3.292$, which is very close 
to the Tonks-Girardeau limit $\pi^2/3$. We have also considered the 
iterated root approximants for the problem of the diluted Fermi 
gas of Sec. 3.3. Extrapolating the ground-state energy from small 
$g\ra\infty$ to the strong-coupling limit $g\ra\infty$ for 
$R^*_4(g)$ provides an accuracy within the maximal error of 
order $10\%$.

\subsection{Odd factor approximants}

In the definition of the factor approximants of odd orders, there 
is a necessity of prescribing the value of one of the parameters, 
say $A_1$, in the general form (3). According to the scaling arguments 
[47,48], this parameter can be set to one. Here we suggest one more 
variant of setting this parameter by defining it as $a_1/a_0$.

Keeping in mind the same problem of extrapolating the ground-state  
energy of the anharmonic oscillator with Hamiltonian (26), we need 
to compare the accuracy of the corresponding factor approximants. The 
standard form is
\be
\label{36}
f_5^*(g) = \frac{1}{2} (1 + g)^{n_1} (1 + A_2 g)^{n_2} 
(1 + A_3 g)^{n_3} \; ,
\ee
where
$$
   A_2 = 21.86082, \qquad A_3 = 8.48018,
$$
$$
   n_1 = 0.27622, \qquad n_2 = 7.16531 \times 10^{-3},
\qquad  n_3 = 0.12584 \; .
$$
In the strong-coupling limit, this gives
\be
\label{37}
f_5^*(g) \simeq 0.669 g^{0.409} \qquad (g\ra \infty) \; .
\ee

Another variant of the factor approximant reads as
\be
\label{38}
f_5^{**}(g) = \frac{1}{2} \left ( 1 +
\frac{a_1}{a_0} \; g \right )^{n_1} (1 + A_2 g)^{n_2}
(1 + A_3 g)^{n_3} \; ,
\ee
with the parameters
$$
   A_2 = 22.16875, \qquad A_3 = 8.83021,
$$
$$
   n_1 = 0.2212, \qquad n_2 = 6.55016 \times 10^{-3}, 
\qquad n_3 = 0.11585.
$$
Now, the strong-coupling limit becomes
\be
\label{39}
f_5^{**}(g) = 0.718 g^{0.344} \qquad (g \ra \infty) \; .
\ee
As is seen, the first form (36) extrapolates better the amplitude, 
while the second variant gives a better extrapolation of the power.
Generally, there is no appriori preference for choosing this or that 
form, which, actually, is in agreement with the scaling arguments 
[47,48].

\subsection{Weighted factor-root approximants}

One more possibility is to construct the weighted approximants 
defined as the linear combination of the factor and root approximants 
of the form
\be
\label{40}
W_k^*(g) = \lbd f_k^*(g) + ( 1 -\lbd) R_k^*(g) \; ,
\ee
with all parameters determined from the accuracy-through-order
matching. The accuracy of such weighted approximants can be
essentially improved. A more general way of constructing average
values for an ensemble of approximants was considered in Ref. [66]

\section{Problem of extrapolation to infinity}

The self-similar factor approximants (3) extrapolate the asymptotic
series (2), valid for small $g\ra 0$, to the region of finite $g>0$.
These approximants, as has been shown by numerous examples [45--49],
provide for finite, even rather large, $g$ very good approximations,
essentially more accurate than Pad\'e approximants.

But the problem, we face now, is to extrapolate series (2) not simply
to finite or large $g$, but to find the limit $f(\infty)$ at
$g\ra\infty$, of course, assuming that this limit exists, so that
\be
\label{41}
f(\infty) = \lim_{g\ra\infty} f(g) = const.
\ee
By requiring that the limit
\be
\label{42}
f_k^*(\infty) = \lim_{g\ra\infty} f_k^*(g) = const
\ee
would also exist, in view of Eq. (3), we come to the condition
\be
\label{43}
\lim_{g\ra\infty} f_0(g) \prod_{i=1}^{N_k} \;
(A_i g)^{n_i} = const \;.
\ee
Let, for concreteness, the behavior of $f_0(g)$ at large $g$ be
$f_0(g) \simeq A g^\al \qquad (g\ra\infty )$, as in Eq. (13).
Then, for condition (43) to hold, it is necessary and sufficient
that
\be
\label{44}
\al + \sum_{i=1}^{N_k} \; n_i = 0 \; .
\ee
Therefore the value of approximant (3), given by
\be
\label{45}
f_k^*(\infty) = A \prod_{i=1}^{N_k} \; A_i^{n_i} \; ,
\ee
provides the approximation for the sought limit $f(\infty)$.

This method of defining the limit $f_k^*(\infty)$ at $g\ra\infty$ by
imposing the restriction (44) on the powers of the approximant (3)
is very simple. However, as has been analyzed in Refs. [45,46], its
accuracy is not high. In the following section, we suggest another
method, whose high accuracy will be illustrated by calculating the
pressure of fluctuating membranes.

\section{Method of variable transformation}

Instead of considering the limit $f(\infty)$ at $g\ra\infty$, it is
convenient to make the change of variables
\be
\label{46}
g = g(z) \; , \qquad z = z(g) \; ,
\ee
such that
\be
\label{47}
\lim_{g\ra\infty} \; z(g) = 1 \; .
\ee
Then, for the function
\be
\label{48}
F(z) \equiv f (g(z)) \; ,
\ee
the sought limit is given by
\be
\label{49}
F(1) = f(\infty)
\ee
at $z=1$.

With the change of variables (46), the series (2) become
\be
\label{50}
f_k(g(z)) \cong F_k(z) \qquad (z\ra 0) \; ,
\ee
where
\be
\label{51}
F_k(z) = F_0(z) \sum_{n=0}^k b_n z^n
\ee
is an expansion in powers of $z$ up to the $k$-th order, with the
coefficients $b_0=1$ and $b_n=b_n(a_1,a_2,\ldots,a_k)$ defined 
through the coefficients $a_n$ of series (2). Then, constructing 
for sum (51) the factor approximant $F_k^*(z)$, we obtain the 
approximation $F_k^*(1)$ for the sought limit $f(\infty)$.

In order to specify transformation (46), we assume the following 
natural properties. According to condition (47), it should be that 
$z\ra 1$ as $g\ra\infty$. If the asymptotic behavior of the sought 
function is
\be
\label{52}
f(g) \simeq f(\infty) \left ( 1 + \frac{C_1}{g^\om} \right )
\qquad (g\ra\infty) \; ,
\ee
we require that transform (48) be
\be
\label{53}
F(z) \simeq F(1) [ 1 + C_2(1-z) ] \qquad (z\ra 1) \; ,
\ee
where $C_1$ and $C_2$ are constants and $\om>0$. And, in agreement 
with Eqs. (50) and (51), we assume that $z\ra 0$ as $g\ra 0$, so 
that
\be
\label{54}
g(z) \simeq \lbd z \qquad (z\ra 0) \; ,
\ee
with a scaling parameter $\lbd>0$.

Comparing Eqs. (52) and (53) requires that
\be
\label{55}
\frac{C_1}{g^\om} \simeq C_2 (1 -z) \qquad (z\ra 1) \; ,
\ee
where we take into account that $f(\infty)=F(1)$. This yields
\be
\label{56}
g(z) \simeq \frac{C_3}{(1-z)^{1/\om} } \qquad (z\ra 1) \; ,
\ee
where $C_3\equiv(C_1/C_2)^{1/\om}$. The interpolation between 
limits (54) and (56) can be done by using the self-similar factor 
approximants [45--50], which results in the form
\be
\label{57}
g(z) = \frac{\lbd z}{(1-z)^{1/\om} } \; .
\ee
Thus, we obtain an explicit expression for the change of variables
(46). For simplicity, we set in what follows the scaling parameter
$\lbd=1$.

But we need yet to define the exponent $\om$. For this purpose, we
introduce the function
\be
\label{58}
\bt(g) \equiv \frac{d\ln f(g)}{d\ln g} =
\frac{g}{f(g)} \; \frac{df(g)}{dg} \; .
\ee
It is easy to notice that, if $f(g)$ enjoys the asymptotic behavior
(52), then the function (58) behaves as
\be
\label{59}
\bt(g) \simeq - \; \om C_1 g^{-\om} \qquad (g\ra\infty) \; .
\ee
Therefore, the exponent $\om$ is defined as
\be
\label{60}
\om = - \lim_{g\ra\infty} \; \frac{\ln|\bt(g)|}{\ln g} \; .
\ee
This procedure is to be accomplished in each approximation order.
That is, for the given $f_k(g)$, defined in Eq. (2), we write
\be
\label{61}
\bt_k(g) = \frac{d\ln f_k(g)}{d\ln g}
\ee
and expand this in powers of $g$, getting
\be
\label{62}
\bt_k(g) = \bt_0(g) \sum_{n=0}^k \; c_n g^n \; ,
\ee
with the coefficients $c_0=1$ and $c_n=c_n(a_1,a_2,\ldots,a_k)$
prescribed by the coefficients $a_n$. Then, we construct the factor
approximant
\be
\label{63}
\bt_k^*(g) = \bt_0(g) \prod_{i=1}^{N_k} \; (1 +D_i g)^{m_i}
\ee
for series (62). From here, considering the limit $g\ra\infty$, 
we get
\be
\label{64}
\om_k = -\lim_{g\ra\infty} \;
\frac{\ln|\bt_k^*(g)|}{\ln g} \; ,
\ee
in analogy with Eq. (60).

Let, for example, the first factor in Eq. (63) behave as
\be
\label{65}
\bt_0(g) \simeq B g^\gm \qquad (g\ra\infty) \; .
\ee
Then approximant (63), at large $g\ra\infty$, is
\be
\label{66}
\bt_k(g) \simeq B g^\gm \prod_{i=1}^{N_k} \;
(D_i g)^{m_i} \qquad (g\ra\infty) \; .
\ee
As a result, Eq. (64) gives
\be
\label{67}
\om_k = - \left ( \gm + \sum_{i=1}^{N_k} m_i \right ) \; .
\ee

To summarize, the calculational scheme is as follows. For a given 
series (2), we find the function (62) and construct its factor 
approximant (63), which provides us with the exponent (67):
\be
\label{68}
f_k(g) \; \ra\; \bt_k(g) \; \ra \; \bt_k^*(g)\; \ra\; \om_k \; .
\ee
Then, making the change of the variable
\be
\label{69}
g = \frac{z}{(1-z)^{1/\om_k} } \; ,
\ee
from $f_k(g)$, we obtain expression (51). Constructing for the 
latter the factor approximant $F_k^*(z)$, we come to the value 
$F_k^*(1)$ approximating the sought limit $f(\infty)$:
\be
\label{70}
F_k(z) \; \ra \; F_k^*(z) \; \ra \; F_k^*(1) \; .
\ee

When constructing the self-similar approximant $F_k^*(z)$, we 
need to solve the system of equations (9). Sometimes, though rare, 
it may happen that Eqs. (9), for some $k$, have no solutions which 
would yield real self-similar factor approximants. In that case, for 
$F_k^*(z)$ we take the arithmetic average of its neighbors 
$F_k^*=(F_{k-1}^*+F_{k+1}^*)/2$. In practical calculations, one  
always deals with the approximation orders $k=1,2,\ldots,K$ up to 
a finite maximal order $K$. Then the final answer for the set of  
$F_k^*$ is given by the average of two last terms 
$(F_K^*+F_{K-1}^*)/2$. The scheme, formulated in the present Section, 
will be applied to studying the properties of fluctuating membranes 
in the following Sections.

\section{Energy of one-dimensional membrane}

A cartoon of a membrane is a one-dimensional string oscillating 
between two rigid walls. This model, to our knowledge, was suggested 
by Edwards [67] and later considered in many articles [2--6,68]. It 
has been shown that calculating the free energy of the string is 
equivalent to finding the ground-state energy of a quantum particle  
in a one-dimensional box. Replacing the rigid walls by a soft potential, 
characterized by a finite stiffness $g$, and employing perturbation 
theory with respect to $g$ yields the series
\be
\label{71}
E_k(g) = \frac{\pi^2}{8g^2} \; \sum_{n=0}^k a_n g^n
\ee
for the particle ground-state energy $E(g)$, with the coefficients
$$
a_0 = 1 \; , \qquad a_1 = \frac{1}{4} \; , \qquad
a_2 = \frac{1}{32} \; , \qquad a_3 = \frac{1}{512} \; ,
$$
$$
a_4 = 0 \; , \qquad a_5 = -\;\frac{1}{131072} \; , \qquad
a_6 = 0 \; , \qquad a_7 = \frac{1}{16777216} \; ,
$$
$$
a_8 = 0 \; , \qquad a_9 = -\; \frac{5}{8589934592} \; ,
$$
and so on.

The series (71) are obtained for the asymptotically small $g\ra 0$.
But, in order to pass to the case of hard walls, one has to consider
the limit $g\ra\infty$, with $E(\infty)$ being the sought value.
Fortunately, the one-dimensional case allows for an explicit solution
[67--69] giving
\be
\label{72}
E(g) = \frac{\pi^2}{8g^2} \left ( 1 + \frac{g^2}{32} +
\frac{g}{4} \; \sqrt{ 1 + \frac{g^2}{64} } \right ) \; ,
\ee
from where
\be
\label{73}
E(\infty) = \frac{\pi^2}{128} = 0.077106 \; .
\ee
This makes it possible to evaluate the accuracy of the self-similar
approximants $E_k^*(\infty)$ with respect to the exact limit (73).

We also wish to compare the accuracy of the two methods described 
above. First, we use the direct method by imposing restriction (44).  
To this end, we construct the factor approximants
\be
\label{74}
E_k^*(g) = \frac{\pi^2}{8g^2} \; \prod_{i=1}^{N_k} \;
( 1 + A_i g)^{n_i}
\ee
for series (71). Imposing the power restriction condition (44), we
have
\be
\label{75}
\sum_{i=1}^{N_k} n_i = 2 \; .
\ee
So that the sought limit is given by
\be
\label{76}
E_k^*(\infty) = \frac{\pi^2}{8}\; \prod_{i=1}^{N_k} A_i^{n_i} \; ,
\ee
according to Eq. (45). The accuracy of approximants (76) is 
characterized by the percentage errors
\be
\label{77}
\ep(E_k^*) \equiv \frac{E_k^*(\infty) - E(\infty)}{E(\infty)} \cdot
100\%
\ee
with respect to the exact value (73).

Another way is to follow the method of Sec. 6. Then we find the 
function (61), with expansion (62), which reads as
\be
\label{78}
\bt_k(g) = - 2 \sum_{n=0}^k c_n g^n \; ,
\ee
with the coefficients
$$
c_0 = 1 \; , \qquad c_1 = -\;\frac{1}{8} \; , \qquad
c_2 = 0 \; , \qquad c_3 = \frac{1}{1024} \; ,
$$
$$
c_4 = 0 \; , \qquad c_5 = -\;\frac{3}{262144} \; , \qquad
c_6 = 0 \; , \qquad c_7 = \frac{5}{33554432} \; ,
$$
$$
c_8 = 0 \; , \qquad c_9 = -\; \frac{35}{17179869184} \; ,
$$
etc.

Constructing the factor approximants (63) for series (78), we find
\be
\label{79}
\bt_k^*(g) = \frac{2g}{\sqrt{64+g^2} }\; - \; 2
\ee
for all $k\geq 4$. Therefore, the exponent (34) is
\be
\label{80}
\om_k= 2 \qquad (k\geq 4) \; .
\ee
Hence, transformation (69) becomes
\be
\label{81}
g = \frac{z}{\sqrt{1-z}} \; .
\ee
Using $g=g(z)$, given by Eq. (81), we get
\be
\label{82}
F(z) \equiv E(g(z)) \; ,
\ee
similarly to Eq. (48). And the related series (81) acquire the form
\be
\label{83}
F_k(z) = \frac{\pi^2}{8z^2} \; \sum_{n=0}^k b_n z^n \; ,
\ee
with the coefficients
$$
b_0 = 1 \; , \qquad b_1 = -\;\frac{3}{4} \; , \qquad
b_2 = -\;\frac{3}{32} \; , \qquad b_3 = -\;\frac{15}{512} \; ,
$$
$$
b_4 = -\; \frac{15}{1024} \; , \qquad b_5 = -\;\frac{1185}{131072} \; ,
\qquad b_6 = -\; \frac{1635}{262144} \; ,
$$
$$
b_7 = -\;\frac{77295}{16777216} \; , \qquad
b_8 = -\; \frac{119595}{33554432} \; , \qquad
b_9 = -\; \frac{24489285}{8589934592} \; ,
$$
and so on.

Then, for series (83), we find the factor approximants
\be
\label{84}
F_k^*(z) = \frac{\pi^2}{8z^2} \; \prod_{i=1}^{N_k} \;
( 1 + B_i z)^{n_i} \; ,
\ee
whose values $F_k^*(1)$ approximate the sought limit $E(\infty)$.
The accuracy of $F_k^*(1)$ is characterized by the percentage errors
\be
\label{85}
\ep(F_k^*) \equiv \frac{F_k^*(1) - E(\infty)}{E(\infty)} \cdot
100\% \; .
\ee

The results of our calculations for the factor approximants
$E_k^*(\infty)$ and $F_k^*(1)$ are presented in Table 1, together
with their errors (77) and (85). As is seen, the method of variable
transformation of Sec. 6 is two orders more accurate than the method
of power restriction employing restriction (44). The final answer,
given by the former method, is $E^*(\infty)=0.0771$, deviating only
by $0.01\%$ from the exact value (73).

\section{Pressure of fluctuating membrane}

An important class of membranes is formed by those membranes whose
constituent molecules can freely move within them. Such membranes
are called fluid. The thermal fluctuations of these membranes, at a
temperature $T$, are controlled by their bending rigidity $\kappa$.
When modeling these membranes, one usually considers them as having
a finite length $L$ and an area $A\ra\infty$. In order to describe
their properties, one, first, assumes that a membrane is located
between the walls of a finite stiffness $g$. This allows one to
resort to perturbation theory in powers of $g$. But to return to the
case of hard walls, one needs to consider the limit $g\ra\infty$,
which requires to invoke a resummation procedure.

It is convenient to introduce the dimensionless pressure $p(g)$ of a
fluctuating membrane, connected with the dimensional pressure $P(g)$
through the relation
\be
\label{86}
p(g) \equiv \frac{\kappa L^3}{8T^2}\; P(g) \; .
\ee
The asymptotic behavior of this function, at small $g\ra 0$, is
represented by the series
\be
\label{87}
p_k(g) = \frac{\pi^2}{8g^2} \; \sum_{n=0}^k a_n g^n \; .
\ee
The coefficients of the perturbation series (87) are known only up to
the sixth order [31], being
$$
a_0=1\; , \qquad a_1 = \frac{1}{4} \; ,
\qquad  a_2 = \frac{1}{32}\; , \qquad
a_3 = 2.176347\times 10^{-3} \; ,
$$
$$
a_4 = 0.552721\times 10^{-4}\; , \qquad
a_5 = -0.721482\times 10^{-5} \; , \qquad
a_6 = -1.777848\times 10^{-6} \; .
$$
We may notice that, up to the second order, the coefficients $a_n$
in pressure (87) are the same as $a_n$ in the ground-state energy 
(71). The pressure of the membrane, located between hard walls, is 
given by the limit
$$
p(\infty) = \lim_{g\ra\infty} p(g) \; .
$$
We shall again find this limit by two methods, by the method 
of the power restriction (44) and the method of the variable 
transformation of Sec. 6.

In the direct method of power restriction (44), we find the 
factor approximants $p_k^*(\infty)$ corresponding to series (87). 
The approximants $p_5^*$ and $p_6^*$ cannot be defined by this way. 
And other approximants are
$$
p_1^*(\infty) = 0.0193\; , \qquad p_2^*(\infty) = 0.0232\; ,
\qquad p_3^*(\infty) = 0.3120\; , \qquad  p_4^*(\infty) = 0.2880\; .
$$
The most accurate Monte Carlo calculations for the membrane pressure
have been accomplished by Gompper and Kroll [33] giving
\be
\label{88}
p_{MC} = 0.0798 \pm 0.0003 \; .
\ee
As we see, the accuracy of $p_k^*(\infty)$, compared to the Monte 
Carlo value (88), is rather bad.

Now we pass to the more elaborated method of Sec. 6, based on 
the change of variables prescribed by Eq. (69). The correct choice 
of the exponent $\om_k$ is very important for achieving a good 
accuracy of the sought limit $F_k^*(1)$. This exponent is expressed 
by Eq. (64) through the function (63), for which we take the even 
factor approximants completely defined in Sec. 2. The series (62), 
for the considered case, has the same form as in Eq. (78), but with 
the coefficients
$$
c_0=1\; , \qquad c_1 = -\;\frac{1}{8} \; ,
\qquad  c_2 = 0 \; , \qquad
c_3 = 0.64173\times 10^{-3} \; ,
$$
$$
c_4 = 0.10668\times 10^{-5}\; , \qquad
c_5 = 0.46253\times 10^{-5} \; , \qquad
c_6 = 0.18454\times 10^{-5} \; .
$$
Again, we may notice that the coefficients $c_0$, $c_1$, and 
$c_2$ for the function $\bt_k(g)$, in the case of the membrane, 
are the same as for series (78) in the case of the string. 
Constructing the even-order factor approximants $\bt_k^*(g)$ 
and substituting these into Eq. (64), we find that the sole real 
exponent $\om_k$ is given by the fourth-order approximant. Thus, 
we are left with the exponent
\be
\label{89}
\om_k= 1.927 \qquad (k\geq 4) \; .
\ee
Accomplishing the change of variables (69), we find series (51), 
for which we construct the factor approximants $F_k^*(z)$. Taking 
the limit $z\ra 1$, we obtain
$$
F_4^*(1) = 0.0906\; , \qquad F_5^*(1) = 0.0898\; , \qquad
F_6^*(1) = 0.0747\; .
$$
Averaging the last two values, we get our final result for the
pressure (86) of the fluctuating membrane: \be \label{90} p(\infty)
= 0.0823 \; . \ee This value is very close to the result [31] of
Kastening $p(\infty)=0.0821$, though it is $3\%$ higher than the
Monte Carlo value (88) of Gompper and Kroll [33]. The achieved
accuracy is quite good, especially keeping in mind that the method
of self-similar factor approximants is much simpler than the
numerical method used by Kastening [31] and several orders simpler
than the Monte Carlo simulations [33].

\section{Conclusion}

We have suggested several modifications for constructing 
self-similar approximants in the frame of the self-similar 
approximation theory. Two main problems are considered, the 
problem of interpolation and extrapolation of asymptotic series. 
The suggested methods are illustrated by examples typical of 
chemical physics and quantum chemistry.

A special attention is payed to the problem of defining the value 
of a function at infinity from its expansion at asymptotically 
small variables. We have designed a new way for constructing the
self-similar factor approximants, so that to derive an accurate
extrapolation $f(\infty)$ for a function $f(g)$ in the limit of
large $g\ra\infty$, when only the asymptotic series $f_k(g)$ at
small $g\ra 0$ are available. We have analyzed and compared two
variants of the extrapolation. One of them involves a restriction 
on the powers of the constructed factor approximants, given by 
Eq.(44). This variant, however, is not sufficiently accurate. The 
latter is caused by the fact that the self-similar factor approximants 
are the most accurate when they are completely defined, by the 
re-expansion procedure, through the coefficients of the asymptotic 
series (2). But imposing additional constraints disturbs the 
self-consistency of the procedure and worsens the accuracy.

The variant of Sec. 6, based on the variable transformation, is
essentially more accurate. This is because it does not involve 
a restriction on powers. Vice versa, it takes into account the 
additional information on the behavior of the function $f(g)$ 
when approaching the limit $f(\infty)$. The prescribed change 
of the variable, not merely tells that $f(\infty)$ is finite, 
but also describes how $f(g)$ approaches this limit. The accuracy 
of the method is illustrated by calculating the pressure of 
fluctuating fluid strings and membranes.

In order to concisely summarize the ideas of the most accurate 
method, let us briefly delineate its main steps. Suppose we aim 
at finding the limit $f(\infty)$ of a function $f(g)$, as 
$g\ra\infty$. But what is known for us is only the approximate 
behavior of the function at asymptotically small $g\ra\infty$, 
where it is approximated by the series $f_k(g)$, and that 
$f(g)\ra const$, as $g\ra\infty$. For a given $f_k(g)$, we define 
the function $\bt_k(g)$ through Eqs. (61) and (62). Then we 
construct the factor approximants $\bt_k^*(g)$, as in Eq. (63), 
and define the exponent $\om_k$ by Eq. (64). According to Eq. (69), 
we make the transformation
$$
g_k(z) = \frac{z}{(1-z)^{1/\om_k} } \; ,
$$and, as in Eq. (50), introduce $F_k(z)=f_k(g_k(z))$. 
Constructing the factor approximants $F_k^*(z)$ and taking the 
limit $z\ra 1$, we obtain the values $F_k^*(1)$ approximating 
the sought-function limit $f(\infty)$. Schematically, all this 
procedure is represented as the sequence of the following steps:
$$
f_k(g) \; \ra \; \bt_k(g) \; \ra \; \bt_k^*(g) \; \ra \; \om_k \;
\ra \; g_k(z) \; \ra
$$
$$ \ra \; F_k(z) \; \ra \; F_k^*(z) \; \ra \;
F^*_k(1) \; \ra \; f(\infty) \; .
$$

We have illustrated the above approach by calculating the 
pressure of fluid fluctuating membranes. The latter form a 
rather widespread important class of membranes studied in 
biology and chemistry [2--6,70]. The asymptotic series for 
the pressure were derived from Helfrich model [5]. The developed 
methods can be applied to other systems, where one needs to 
extrapolate the sought function from the region of asymptotically 
small variables to their finite values. Moreover, the suggested 
methods make it even possible to find, with a good accuracy, the 
limit of the function at infinity. The advantage of the developed 
methods is their simplicity and high accuracy.

\newpage

\begin{center}

{\large{\bf Table 1}}

\vskip 1cm

\begin{tabular}{|r|r|r|r|r|} \hline
 $k$ & $E_k^*(\infty)$ & $\%$ & $F_k^*(1)$ & $\%$   \\ \hline
 3   & 0.1500  &  94.1 & 0.0593 & -23.1  \\
 4   & 0.1370  &  78.1 & 0.0935 &  21.3  \\
 5   & 0.0526  & -31.8 & 0.0926 &  20.0    \\
 6   & 0.0550  & -28.7 & 0.0829 &  7.52 \\
 7   & 0.1030  &  33.4 & 0.0732 & -5.01   \\
 8   & 0.0993  &  28.8 & 0.0805 &  4.36  \\
 9   & 0.0620  & -19.6 & 0.0803 &  4.17  \\
10   & 0.0636  & -17.5 & 0.0783 &  1.49 \\
11   & 0.0926  &  20.1 & 0.0762 & -1.24  \\
12   & 0.0906  &  17.5 & 0.0780 &  1.13 \\
13   & 0.0662  & -14.1 & 0.0779 &  1.08  \\
14   & 0.0674  & -12.6 & 0.0774 &  0.37 \\
15   & 0.0882  &  14.4 & 0.0768 & -0.34   \\  \hline
$E^*(\infty)$ & 0.0778 & 0.91 & 0.0771 & -0.008 \\ \hline
\end{tabular}

\end{center}

\vskip 5mm

\parindent=0pt

The factor approximants $E_k^*(\infty)$ and $F_k^*(1)$, together
with their percentage errors, approximating the ground-state energy
$E(\infty)$.

\end{document}